\documentclass[10pt,aps,prd,reprint,nofootinbib,floats,floatfix,amsfonts,amssymb,amsmath,preprintnumbers,notitlepage,superscriptaddress,longbibliography]{revtex4-1}
\usepackage{graphicx}
\usepackage[utf8]{inputenc}
\usepackage[british]{babel}
\usepackage{isomath}
\usepackage[protrusion=true,tracking=true,kerning=true,spacing=true,final,babel=true]{microtype}
\usepackage{physics}
\usepackage{xcolor}

\begin{document}

\begin{flushright}
KCL-PH-TH-2021-68
\end{flushright}

\title{
Footprints of population III stars in the gravitational-wave background}
\author{Katarina Martinovic}
\affiliation{Theoretical Particle Physics and Cosmology Group, \, Physics \, Department, \\ King's College London, \, University \, of London, \, Strand, \, London \, WC2R \, 2LS, \, UK}
\author{Carole P\'{e}rigois}
\affiliation {LAPP, CNRS, 9 Chemin de Bellevue, 74941 Annecy-le-Vieux, France}
\author{Tania Regimbau}
\affiliation {LAPP, CNRS, 9 Chemin de Bellevue, 74941 Annecy-le-Vieux, France}
\author{Mairi Sakellariadou}
\affiliation{Theoretical Particle Physics and Cosmology Group, \, Physics \, Department, \\ King's College London, \, University \, of London, \, Strand, \, London \, WC2R \, 2LS, \, UK}

\date{September 2021}

\begin{abstract}
    We investigate detection prospects of the gravitational-wave background (GWB) that originates from the merging of compact objects formed by the collapse of population III stars. Younger population I/II stars lead to a GWB in the LIGO/Virgo frequency band at the inspiral phase, while population III stars would likely show up at the later merger and ringdown phases. We show that, using a network of third-generation detectors, we may be able to separate a population I/II signal from a population III one, provided we can subtract individual coalescence events. A detection of a population III GWB could reveal important information, such as the average redshifted total mass.

\end{abstract}

\maketitle

\textbf{Introduction---}
We have witnessed a rapid expansion of gravitational-wave (GW) astrophysics in the last decade due to the success of the Advanced LIGO and Virgo GW detectors \cite{2015AdvancedLIGO, Acernese_2014} at uncovering signals from numerous compact binary coalescences (CBCs) \cite{KAGRA:2013rdx}. The most recent Advanced LIGO/Virgo observing run, O3, presented us with dozens of new merger events and has significantly expanded the stellar graveyard \cite{LIGOScientific:2020ibl}. Despite the increase in the number of detections, we are yet to observe with confidence an event that would suggest that the progenitor compact objects are remnants of the oldest stars in the Universe \cite{LIGOScientific:2020kqk} - the theoretically-postulated population III (pop III) stars (GW190521 could be a potential candidate \cite{Pacucci:2017jbp}). Pop III stars are thought to have formed at high redshifts and as such have low metallicity compared to the more recently formed, population I/II (pop I/II) stars \cite{Bromm:1999du,Bromm:2003vv,Yoshida:2003rw}. These old stars have hitherto evaded sky surveys \cite{deSouza:2011ea,Bowler:2016qen,2019BAAS...51c.449W}, and their detection remains an objective for upcoming experiments, such as the James Webb Space Telescope \cite{Gardner:2006ky}. 

Pop III stars may solve some of the puzzles in black hole formation, as well as help understanding the early epochs of the Universe such as reionisation and galaxy evolution \cite{1984ApJ...277..445C, Yoshida:2003ab, Trenti:2009cj}. Numerical simulations show that these primordial stars could have led to the formation of super-massive black holes at high redshifts \cite{2018MNRAS.475.4104C, Hirano:2015wxa, Heger:2002by}. Mergers of such heavy remnants would appear in the millihertz frequency range explored by future space-based detectors such as LISA. The scope of this study, however, is detection prospects of terrestrial detector networks, and we therefore focus on models that predict a pop III signal in the LIGO/Virgo frequency range. The contribution to the gravitational-wave background (GWB)\footnote{Often referred to as the \textit{stochastic} gravitational-wave background \cite{Rosado:2011kv}.} from a superposition of unresolved pop III-seeded CBCs has been explored in several studies \cite{ Belczynski:2001uc, Suwa:2007du, Inayoshi:2016hco,Belczynski:2016ieo, Ng:2020qpk}.
They show significant deviation of a pop III star signals from a pop I/II stars signal due to different mass and redshift distributions \cite{Ng:2020qpk}. The GWB is comprised of many sources, of astrophysical or cosmological origin, but we expect the CBC signal to be the foreground to all sources \cite{KAGRA:2021kbb}. In this study, for the first time, we consider the possibility of separating pop I/II and pop III GWB contributions. Numerous models suggest the total CBC background is dominated by pop I/II. However, pop III can be uncovered using subtraction techniques and studying the residual backgrounds \cite{Regimbau:2014uia,Sachdev:2020bkk}. As the sensitivity of detectors increases and GW interferometers {\sl see} more individual CBC events, a pop III residual background emerges as the dominant signal over pop I/II residual background. We first study how to detect the GWB from pop III stars, and in the case of a successful detection, we explore subsequent implications, namely information about masses and redshifts of the population.

The paper is organised as follows: in Sec.~II we present the pop III models and their resulting GWB in different detector networks, highlighting characterisation of the residual backgrounds. We then introduce  Bayesian analysis used in Sec.~III, and discuss search filters we consider for pop III stars. Sec.~IV is an implications study in case of a detection of a pop III signal. We use StarTrack (ST) simulation data \cite{Belczynski:2005mr} and apply our detection methods, ultimately showing consistency of our implications analysis with the underlying population. We select the ST data since it is the most recent extensive catalogue of merging binaries from pop I, II and III stars that lead to a GWB in the LIGO/Virgo frequency range \cite{Perigois:2020ymr}.

\textbf{Population III GWB---}
The GWB is defined as the superposition of GWs from all unresolved sources. It is characterised by the dimensionless parameter $\Omega_{\rm GW}$ \cite{Allen:1997ad}, expressed as the ratio of GW energy density per logarithmic frequency bin $\text{d}\rho_{\rm GW}/\text{d}\,\text{ln}(f)$, normalised by the critical energy density of the Universe, $\rho_{\rm c} = (3H_0^2c^2)/(8\pi G)$:
\begin{equation}
    \Omega_{\rm GW}(f) = \frac{1}{\rho_{\rm c}} \frac{\text{d} \rho_{\rm GW}(f)}{\text{d}  \text{ln}(f)},
\end{equation}
with $H_0=67.9$ km s$^{-1}$ Mpc$^{-1}$ \cite{Planck:2018lbu}.

Here we concentrate on the CBC contribution to the GWB, namely from pop I/II stars and the theoretical pop III stars. One can express the quantity $\Omega_{\rm GW}$ in terms of CBC source parameters $\theta$ (masses and spins), as \cite{Perigois:2020ymr} :

\begin{equation}
    \Omega_{\rm GW}(f)=\frac{f}{\rho_c H_0} \int \text{d}\theta  p(\theta) \int_0^{z_{\rm up}(\theta)} \text{d}z \frac{R(z;\theta) \frac{\text{d}E_{\rm GW}(f_s;\theta)}{\text{d}f_s}}{(1+z) E_z(z)},
\label{eq:omega_analyic}
\end{equation}
where $p(\theta)$ is the probability distribution of the source parameters, ${\text{d}E_{\rm GW}}/{\text{d}f_{\rm s}}$ in the energy density emitted by a single source at a redshift $z$ with parameters $\theta$, $f_{\rm s}$ is the emitted frequency in the source frame $f_{\rm s} = f(1+z)$ and  $z_{\rm up}(\theta)$ is the maximal redshift at which a compact binary with parameters $\theta$ can form. The factor $(1+z)$ in the denominator converts the merger rate $R(z,\theta)$ from the source to the detector frame, and $E_z(z)$ accounts for the considered cosmology, i.e. the expansion history of the Universe, $E_z(z) = \sqrt{\Omega_{\rm m}(1+z)^3 +\Omega_\Lambda},$ with $\Omega_{\rm m}=0.31, 
\Omega_\Lambda =0.69$ \cite{Planck:2018lbu}.

For the total population, when all sources are included in the background, the merger rate $R(z;\theta)$ at redshift $z$ for sources with parameters $\theta$ is given in the source frame, per unit of comoving volume and time. It is derived from the star formation rate, corrected by the time delay between the birth of the progenitors and the merger of the compact objects \cite{Barack:2018yly}.
To calculate the residual background when individually detected sources are removed, one has to multiply the total rate by a factor $1-\epsilon(z,\theta)$, where the efficiency $\epsilon(z,\theta)$ is the probability for a source at redshift $z$ with parameters $\theta$ to be detected, integrated over inclination, polarisation and position in the sky (see \cite{Regimbau:2014nxa}).

In the case of binary neutron stars and neutron star-black hole mergers, we only consider the inspiral phase and assume that the emission of GWs stops at the last stable orbit.
For binary black holes, we consider the three different regimes of the coalescence (inspiral, merger and ringdown phase) given by the corresponding phenomological waveforms  \cite{Ajith:2009bn} calculated for circular orbits. The energy density is  \cite{Regimbau:2008nj}:
\begin{align}
 \label{eq:dedf}
   \frac{\text{d}E_{\rm GW}}{\text{d}f_{\rm s}} (f_{\rm s})&=\frac{5 (G \pi)^{2/3} \mathcal{M}_{\rm c}^{5/3} F_\iota }{12} f_{\rm s}^{-1/3}  \\ \nonumber
   & \times  \begin{cases}
             (1+\sum^3_{i=2} \alpha _i\nu^i)^2 & \text{if } f_s<f_{\rm merg}  \\
 f_{\rm s}  w_{\rm m} (1+\sum^2_{i=1} \epsilon_i\nu^i)^2 & \text{if }f_{\rm merg}\leq f_{\rm s}<f_{\rm ring}  \\
 f_{\rm s}^{1/3} w_{\rm r}  \mathcal{L}^2(f_{\rm s},f_{\rm ring},\sigma) & \text{if } f_{\rm ring}\leq f_{\rm s}<f_{\rm cut}
  \end{cases}
\end{align}
where $f_{\rm merg}$, $f_{\rm ring}$ and $f_{\rm cut}$ are the frequencies at the start of merger, start of ringdown and end of emission in the source frame, respectively. The chirp mass of the binary is a combination of the individual masses of the compact objects, $\mathcal{M}_{\rm c}=\frac{(m_1 m_2)^{3/5}}{(m_1+m_2)^{1/5}}$. $\mathcal{L}(f_{\rm s}, f_{\rm ring}, \sigma)$ is the Lorentzian function centered at $f_{\rm ring}$, with width $\sigma$, and $w_m$, $w_r$ are the normalisation constants ensuring the continuity between the three phases. The factors $\epsilon_i$ and $\alpha_i$ and the frequencies $f_{\rm merg}$, $f_{\rm ring}$ and $f_{\rm cut}$ follow from analytical waveforms detailed in \cite{Ajith:2009bn} and depend on the symmetric mass ratio $\eta =(m_1m_2)/(m_1+m_2)^2$ of the progenitors' masses, $m_1$ and $m_2$, and the effective spin of the system $\chi = [(m_1\vec s_1 + m_2\vec s_2)/(m_1+m_2) ]\vec L/L$. 
Once the spectrum of $\Omega_{\rm GW}$ is calculated, we can estimate the corresponding signal-to-noise ratio ($\text{SNR}$) for a given network of $N$ detectors \cite{Romano:2016dpx}: 
\begin{equation}
    \text{SNR} =\frac{3 H_0^2}{10 \pi^2} \sqrt{2T} \left[
\int_0^\infty \text{d}f\>
\sum_{I=1}^N\sum_{J>I}
\frac{\gamma_{IJ}^2(f)\Omega_{\rm GW}^2(f)}{f^6 P_J(f)P_J(f)} \right]^{1/2}\,,
\label{eq:snrCC}
\end{equation}
with $T$ the observational time, $P_I$ and $P_J$ the one-sided power spectral noise densities of detectors $I$ and $J$, and $\gamma_{IJ}$ the normalised isotropic overlap reduction function characterising the distance and the relative orientation between $I$ and $J$ for sources isotropically distributed in the sky.

In this work, we consider the StarTrack model FS1 for pop III \cite{Belczynski:2016ieo}. Assumptions about characteristics of the initial binary pop III stars are discussed in \cite{10.1093/mnras/stv2629}. The FS1 model assumes that pop III stars were formed in large gas clouds with a star formation rate that peaks at redshift $z \sim 12$, while the star formation rate for pop I/II stars peaks at $z \sim 2$ (see Fig.~4 in \cite{Belczynski:2016ieo}).
Even though pop III stars are less abundant than pop I/II, the ST model FS1 considers a rather  optimistic ratio of pop III to pop I/II stars. The corresponding background and its detectability have been calculated \cite{Perigois:2020ymr} using a catalogue of sources rather than the analytical expression in Eq.~\ref{eq:omega_analyic}. The residual background 
catalogue is obtained by subtracting all sources individually detected by the interferometer network. For each source $k$ we calculate the individual SNR $\rho^k$ assuming optimal-matched filtering and uncorrelated gaussian noise in the detectors as follows:
\begin{equation}
    \left( \rho^{k} \right)^2 = \sum_{I=1}^N 4 \int_{f_{i,\min}}^{f_{i,\max}} \frac{\left |\tilde{H}^k \right |^2}{P_I(f)}\ df,
 \label{eq:rho}
\end{equation}
where
\begin{equation}
    \tilde{H}^k=F_{+,I}(f,\Theta^k,\psi^k) \tilde{h}^k_{+}(f)
 +F_{\times,I}(f,\Theta^k,\psi^k) \tilde{h}^k_{\times}(f),
 \end{equation}
with $F_{+,I}$ and $F_{\times,I}$ the antenna factors of detector $I$ for  polarisations $+$ and $\times$ that depend on source inclination $\Theta^k$ and position in the sky $\psi^k$, 
while $\tilde{h}_{+}^k$ and $\tilde{h}_{\times}^k$ are the Fourier transforms of the gravitational waveforms of the source $k$. A residual catalogue is computed by removing all sources with $\rho^k > 12$.

If pop III exists, its signal will be superposed with a pop I/II signal. In the case of a dominant pop I/II signal, the pop III signal will remain hidden underneath it, and one can only place upper limits on the amplitude of the pop III contribution to $\Omega_{\rm GW}(f)$. If, however, a pop III signal is the dominant one, then we could detect deviations from the 2/3 CBC power law, and even getting insight on the mass and redshift distribution of pop III stars. We explore the last scenario by considering two terrestrial networks of third-generation (3G) detectors:
(i) Einstein Telescope (ET) at the  Virgo site, and (ii) ET at the Virgo site with two Cosmic Explorers (CE) at the LIGO Hanford and Livingston sites. 

Estimates of CBC contributions to $\Omega_{\rm GW}$ from ST simulations suggest that pop III signal is lost in the pop I/II foreground. For 2G detector networks -- even by including LIGO-Hanford, LIGO-Livingston, Virgo, LIGO-India, and KAGRA -- pop III is practically invisible and its contribution to the global SNR is negligible, as it is shown in \cite{Perigois:2020ymr}. 
However, 3G detectors such as  ET and CE,  may reveal a pop III background. The future detectors will have unprecedented sensitivity and they will be able to discover a great number of individual CBCs, thereby reducing the GWB originating from unresolved CBCs. For ET+2CE, we uncover pop III after the subtraction of individually resolved merger events. This follows because subtraction methods are less efficient to detect the high redshift and low frequency pop III CBCs. Being more difficult to resolve, binaries from pop III persist, resulting in a large contribution to the residual CBC background in 3G detectors. 

\begin{figure}
    \centering
    \includegraphics[width=6.5cm]{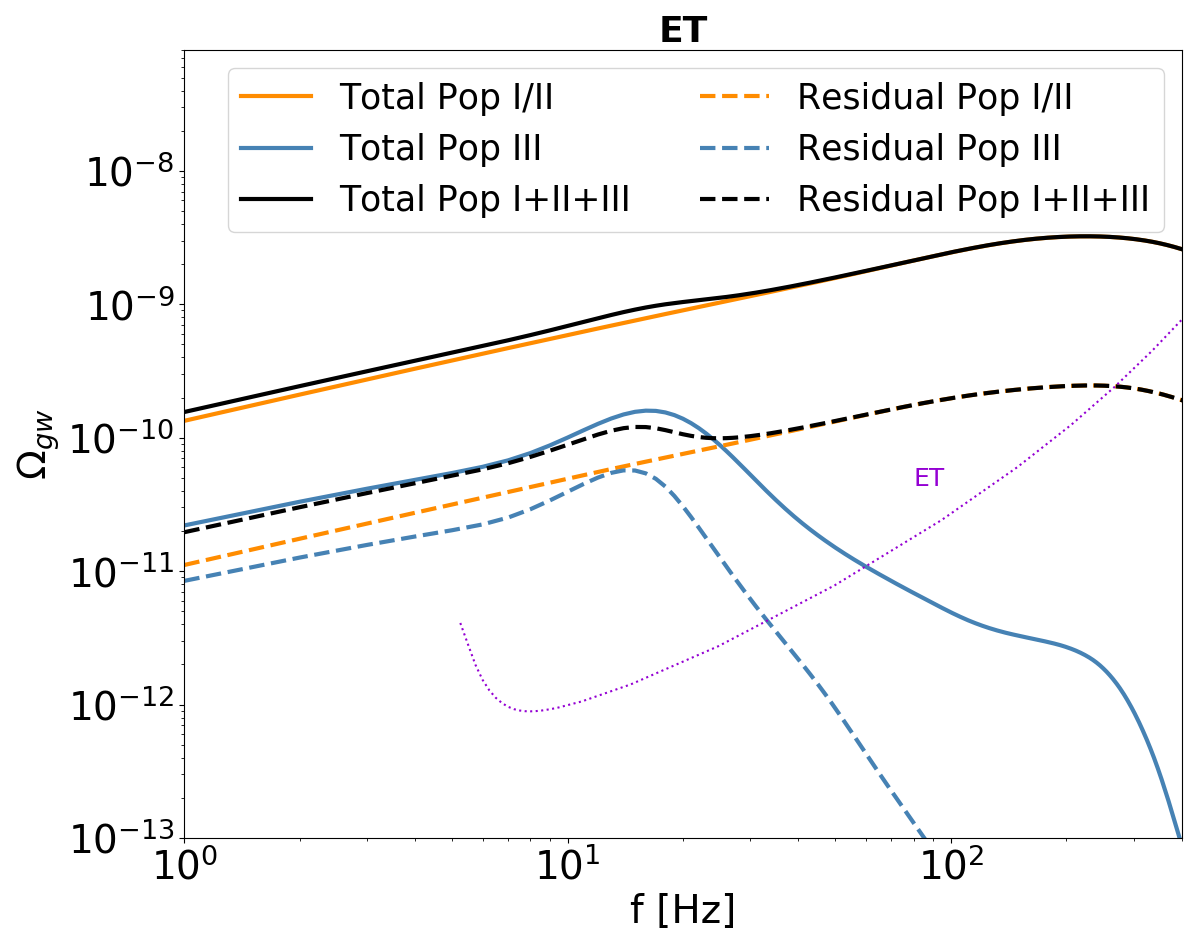}
    \includegraphics[width=6.5cm]{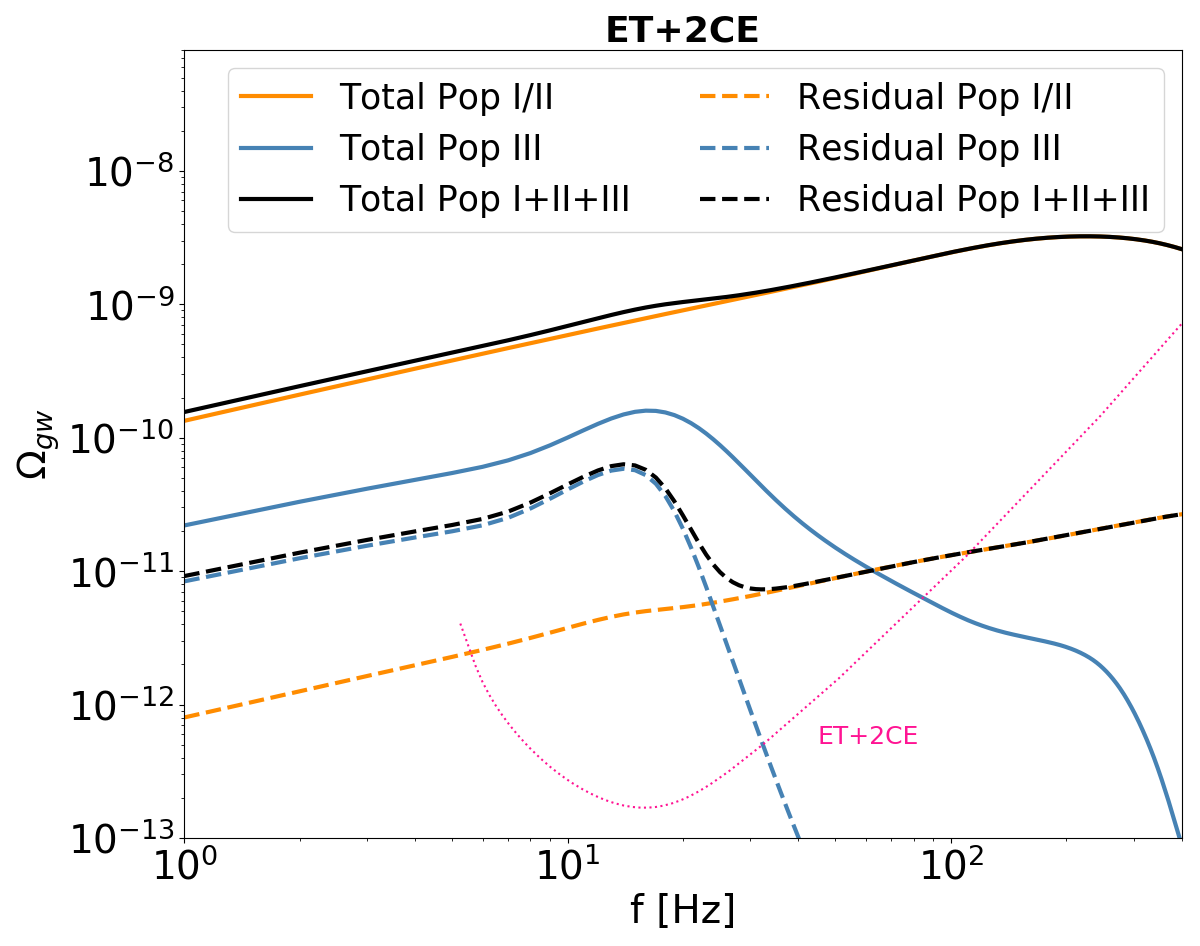}
    \caption{Total and residual GWB of ET (top) and ET+2CE (bottom) detector networks. The pop I/II and pop III contributions are shown in green and red, respectively, with the combined residual signal shown in black.}
    \label{fig:Residual_bkg}
\end{figure}

We compare in Fig.~\ref{fig:Residual_bkg} the total and residual background for the two 3G networks: ET (top) and ET+2CE (bottom). It confirms that the pop III contribution in ET has a very small impact on the combined residual background from pop I/II and pop III, while in ET+2CE the pop III residual background clearly dominates for frequencies below $\sim$ 20Hz. In addition, Fig.~\ref{fig:Residual_bkg} shows a change in the shape of the background: The peak frequency of pop III changes slightly while the slope characterising the end of emission decreases dramatically when we remove individually detected sources. 

\begin{figure}
    \centering
    \includegraphics[width=6.5cm]{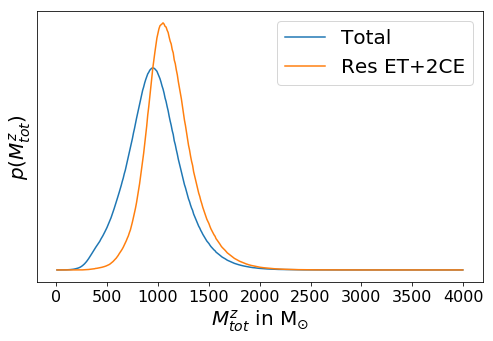}
    \includegraphics[width=7cm]{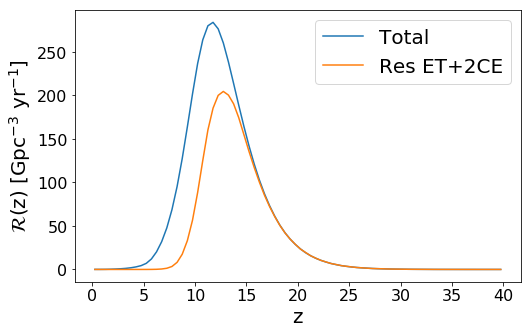}
    \caption{Comparison between the total (blue) and ET+2CE residual (orange) catalogue for redshifted total mass distributions (top) and merger rates (bottom).}
    \label{fig:distrib}
\end{figure}

To demonstrate the impact of subtraction of resolved CBCs on the population, we show in Fig.~\ref{fig:distrib} the probability density of the total redshifted mass, $M_{\rm{tot}}^z=(1+z)(m_1+m_2)$, and the merger rate $R(z)$ as a function of redshift, between the whole catalogue and the residual one for ET+2CE.
Clearly, the sources remaining in the residual catalogue are the ones with the highest redshift, affecting the total corrected mass distribution which is in turn responsible for the changes in the GWB spectrum. We will estimate the ET+2CE residual pop III parameters by filtering the corresponding background and performing a Bayesian analysis.

\textbf{Detection method---}
The stochastic pipeline takes strain data $\tilde{s}_{I,J}$ from detectors, $I, J$, and constructs cross-correlation statistics using optimal filters \cite{Romano:2016dpx}:
\begin{equation}
        \label{eq:methods:bin_by_bin_estimator}
\hat{C}_{IJ}(f) =\frac{2}{T}  \frac{{\rm Re}[\tilde{s}_I^*(f) \tilde{s}_J(f)]}{\gamma_{IJ}(f) S_0(f)},
\end{equation}
with $T$ the duration of the run, and $\gamma_{IJ}(f)$ the normalised overlap reduction function as defined in Eq.~\ref{eq:snrCC}. The estimator is normalised with $S_0(f) = (3 H_0^2)/(2\pi^2f^3)$ leading to $\langle \hat{C}_{IJ}(f) \rangle = \Omega_{\rm GW}(f)$. We assume correlated noise not to be a limiting factor to our detector sensitivity and consider all noise to be gaussian. The variance is 
\begin{equation}
\label{eq:def_sigma}
\sigma_{IJ}^2(f) \approx \frac{1}{2T \Delta f}  \frac{P_I(f) P_J(f)}{\gamma_{IJ}^2(f) S_0^2(f)}.
\end{equation}
Let us construct a gaussian log-likelihood,
\begin{eqnarray}
\label{eq:likelihood}
p(\hat{C}_{IJ}(f | \theta)) &\propto&\exp\left[  -\frac{1}{2} \sum_{IJ}  \left(\frac{\hat{C}_{IJ}(f) - \Omega_{\rm GW}(f|\theta)}{\sigma_{IJ}(f)}\right)^2  \right], \nonumber \\
\end{eqnarray}
where $\Omega_{\rm GW}(f|\theta)$ represents the GWB model with parameters $\theta$. This allows us to estimate the model parameters by finding the best-fit to the cross-correlation data and minimising the likelihood function. Note that we have made the simplifying assumption that the log-likelihood of a detector network is the sum of log-likelihoods of the individual baselines. To compare models and find which ones fit data better, we perform model selection with Bayes factors. Bayes factor, $\mathcal{B}^{\mathcal{M}_1}_{\mathcal{M}_2}$, is defined as the ratio of evidences of model $\mathcal{M}_1$ to model $\mathcal{M}_2$, and if large and positive, demonstrates preference for $\mathcal{M}_1$ over $\mathcal{M}_2$.

Typically, we model a CBC signal as $\Omega_{\rm GW}(f)=\Omega_{\rm ref}~(f/f_{\rm ref})^{2/3}$, with $f_{\rm ref}$ = 25 Hz. This is because the CBCs detected so far have low masses that would lead to an inspiral signal in the low-frequency range. This can be seen in Fig. \ref{fig:Residual_bkg} where the total GWB from pop I/II and III in 3G detectors are presented. The pop III spectrum shows clear deviation from a 2/3 power law, because these further away stars will lead to more redshifted frequencies and therefore be detected in their merger and ringdown phases. We test search filters different from a 2/3 power law to investigate if the deviation from pop I/II signal can be identified in a parameter estimation study. Motivated by the shape of the residual pop III signal in Fig.~\ref{fig:Residual_bkg}, we consider the following filters: 
\begin{itemize}
    \item power law with varying spectral index (PL)
\begin{equation}
    \Omega_{\rm GW}^{\rm PL}(f) = \Omega_{\rm ref}~(f/f_{\rm ref})^{\alpha}
\end{equation}

    \item broken power law (BPL)
\begin{equation}
      \Omega_{\rm GW}^{\rm BPL}(f) =
  \begin{cases}
              \Omega_{\rm peak} (f/f_{\rm peak})^{\alpha_1} & \text{for $f \leq f_{\rm peak}$}, \\
              \Omega_{\rm peak} (f/f_{\rm peak})^{\alpha_2} & \text{for $f > f_{\rm peak}$}.
  \end{cases}
\end{equation}

    \item smooth BPL
\begin{equation}
    \Omega_{\rm GW}^{\rm SBPL}(f) =
              \Omega_{\rm peak}~(f/f_{\rm peak})^{\alpha_1}~ [1+(f/f_{\rm peak})^{\Delta}]^{(\alpha_2-\alpha_1)/\Delta}.
\end{equation}

    \item triple BPL
\begin{equation}
      \Omega_{\rm GW}^{\rm TBPL}(f) =
  \begin{cases}
              \Omega_{\rm peak} (f/f_{\rm peak}^{(1)})^{\alpha_1} & \text{for $f \leq f_{\rm peak}^{(1)}$}, \\
              \Omega_{\rm peak} (f/f_{\rm peak}^{(1)})^{\alpha_2} & \text{for $ f_{\rm peak}^{(1)} < f \leq f_{\rm peak}^{(2)}$}, \\
              k \, \Omega_{\rm peak} (f/f_{\rm peak}^{(2)})^{\alpha_3} & \text{for $f > f_{\rm peak}^{(2)}$},
  \end{cases}
\end{equation}
where $k=(f_{\rm peak}^{(2)} / f_{\rm peak}^{(1)})^{\alpha_2}$ ensures continuity of the piecewise function.
\end{itemize}

The priors for each model's parameters can be found in the Appendix. If any of the filters above are preferred over a 2/3 filter, this could be an indication of the presence of a pop III signal. 

\textbf{Implications---}
In the case of a detection, we examine whether we can constrain the mass/redshift distribution from the optimal search parameters. Following the GWB expression (Eq.~(\ref{eq:omega_analyic})), we see that the parameters impacting the background shape are the redshift-dependent merger rate and the black holes' mass distribution. To understand how these population characteristics relate to model parameters, such as peak frequency and slope, we generate multiple spectra. We make simplifying assumptions about our progenitors by assuming spinless, equal-mass binaries \cite{Gerosa:2020bjb}. We fix the merger rate and vary the intrinsic mass input, observing how the shape of the GWB spectrum changes. The results we find, however, change with a different choice of merger rate, as described in the Appendix. This is because there is a degeneracy between the effects that merger rate and mass distribution have on the GWB \cite{Inayoshi:2016hco}. We thus study the dependence of $\Omega_{\rm GW}$ on the \textit{redshifted} total mass of the population, $M^z_{\rm tot}=(1+z)(m_1+m_2)$, which is related to the merger rate, and find a relationship between the mass and the peak frequency of the spectrum.

We generate GW spectra with a merger rate from ST, varying the redshifted total mass, and we find an agreement (within 10\%) between redshifted ringdown frequency and the peak of the spectrum, see Table~\ref{tab:m_z}. We obtain the same agreement if we use the merger rate from \cite{Inayoshi:2016hco}, suggesting once more that an estimate of the peak frequency can be used to constrain the average redshifted total mass of the population. This relationship, therefore, holds independently of the model used for the evolution of the pop III binaries.

\begin{table}[h!]
    \centering
    \begin{tabular}{c|c|c|c}
        $M_{\rm{tot}}^z$ & $f_{\rm peak}$ & $f^z_{\rm ring}$ & \% difference \\
        \hline 100 & 166.2 & 165.8 & 0.20 \\
        200 & 83.7 & 82.9 & 1.0\\
        300 & 56.5 & 55.3 & 2.1 \\
        400 & 43.1 & 41.4 & 3.9 \\
        500 & 35.2 & 33.2 & 5.7 
    \end{tabular}
    \caption{Variation of the peak of GWB spectra with a change in redshifted total mass. We find agreement between the peak frequency and the redshifted ringdown frrequency.}
    \label{tab:m_z}
\end{table}

\textbf{Results---}
We simulate one year of observation time with the ET+2CE network, taking the CBC background from the ST catalogue. We find the best-fit models to the \textit{residual} GWB that remains after subtracting the individual sources. A model selection study shows preference for other filters over a 2/3 PL, see Table~\ref{tab:BFs}. The models with a broken power law shown in the last 3 rows are clearly favoured over a single power law model. However, we do not observe a great increase in Bayes factor for the smooth and triple BPL over just a BPL. Therefore, we conclude that a BPL filter is sufficient for a pop III GWB search.

\begin{table}[h!]
    \centering
    \begin{tabular}{c|c}
        model, $\mathcal{M}$ & ln $\mathcal{B}^{\mathcal{M}}_{2/3}$ \\
        \hline PL & 29 000\\
        BPL & 46 000 \\
        smooth BPL & 47 000 \\
        triple BPL & 46 000 
    \end{tabular}
    \caption{Log Bayes factor of pop III filters compared to the 2/3 power law filter.}
    \label{tab:BFs}
\end{table}

\begin{figure}
    \centering
    \includegraphics[width=0.4\textwidth]{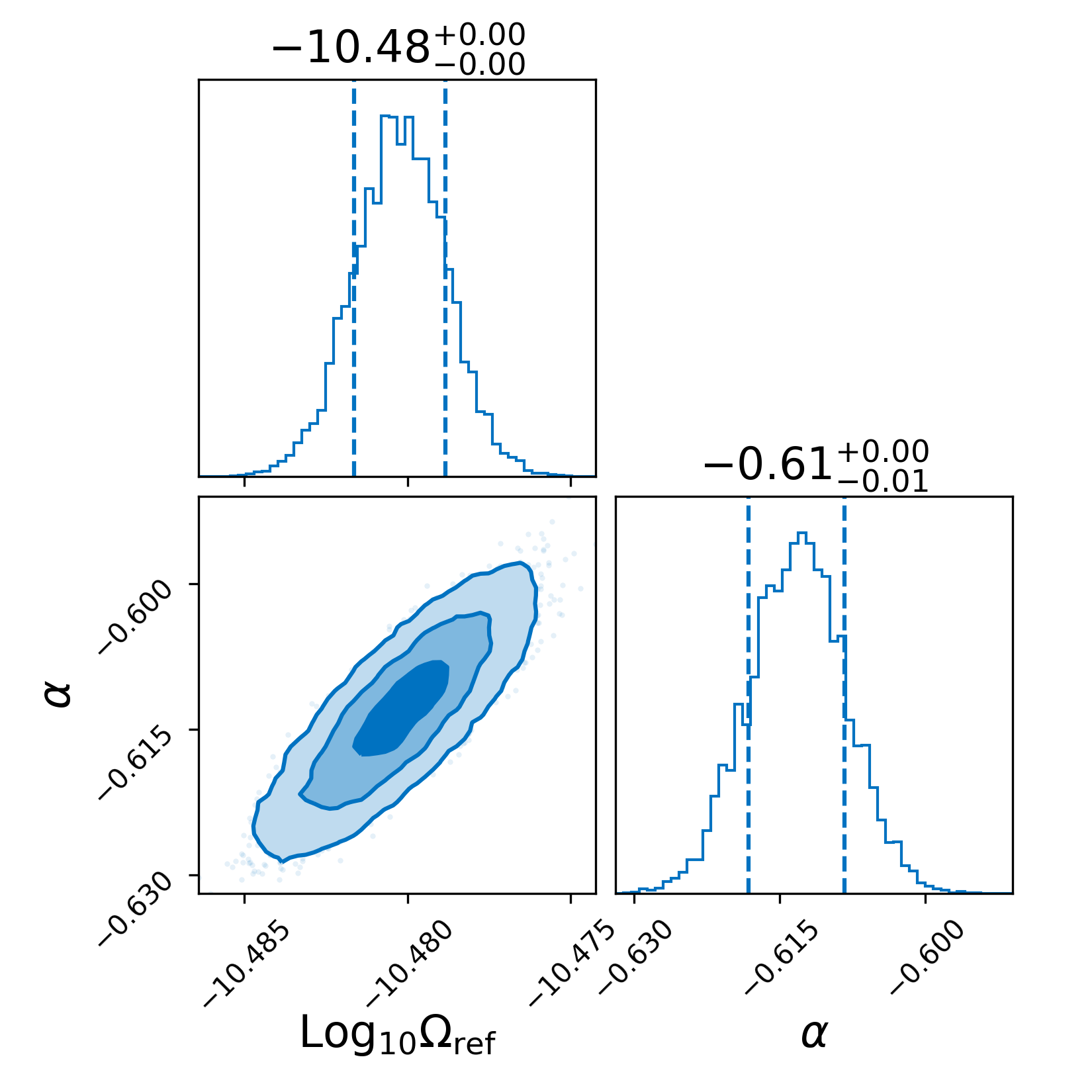}
    \caption{Varying-$\alpha$ PL fit to residual GWB spectrum of pop I+II+III from the ST simulation. We see that the $\alpha$ estimate is not 2/3 which would be expected for pop I/II.}
    \label{fig:corner_pl}
\end{figure}

\begin{figure}
    \centering
    \includegraphics[width=0.5\textwidth]{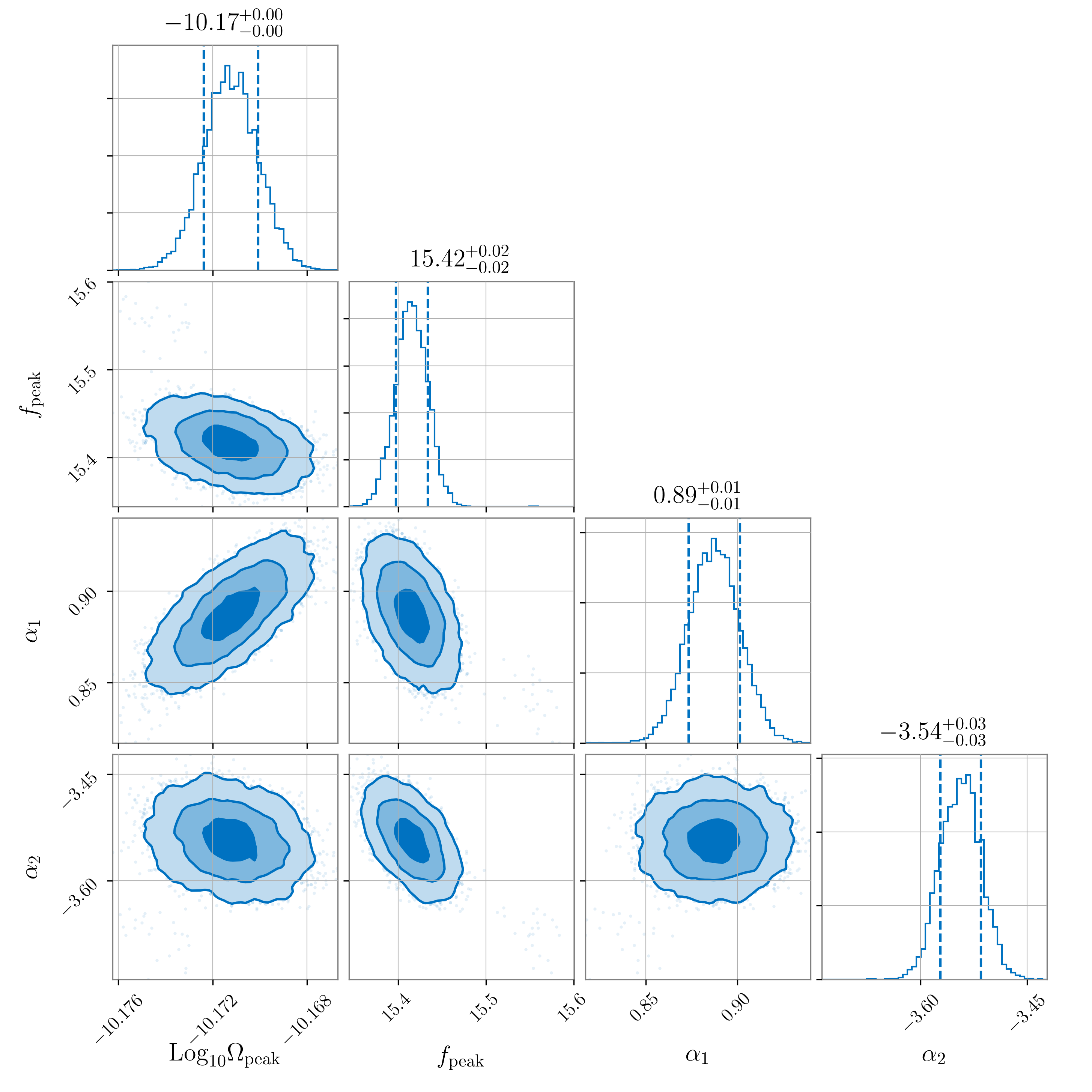}
    \caption{BPL fit to residual GWB spectrum of pop I+II+III from the ST simulation. The peak frequency is estimated to $f_{\rm peak} = 15.4$ Hz.}
    \label{fig:corner_bpl}
\end{figure}

\begin{figure}
    \centering
    \includegraphics[width=0.4\textwidth]{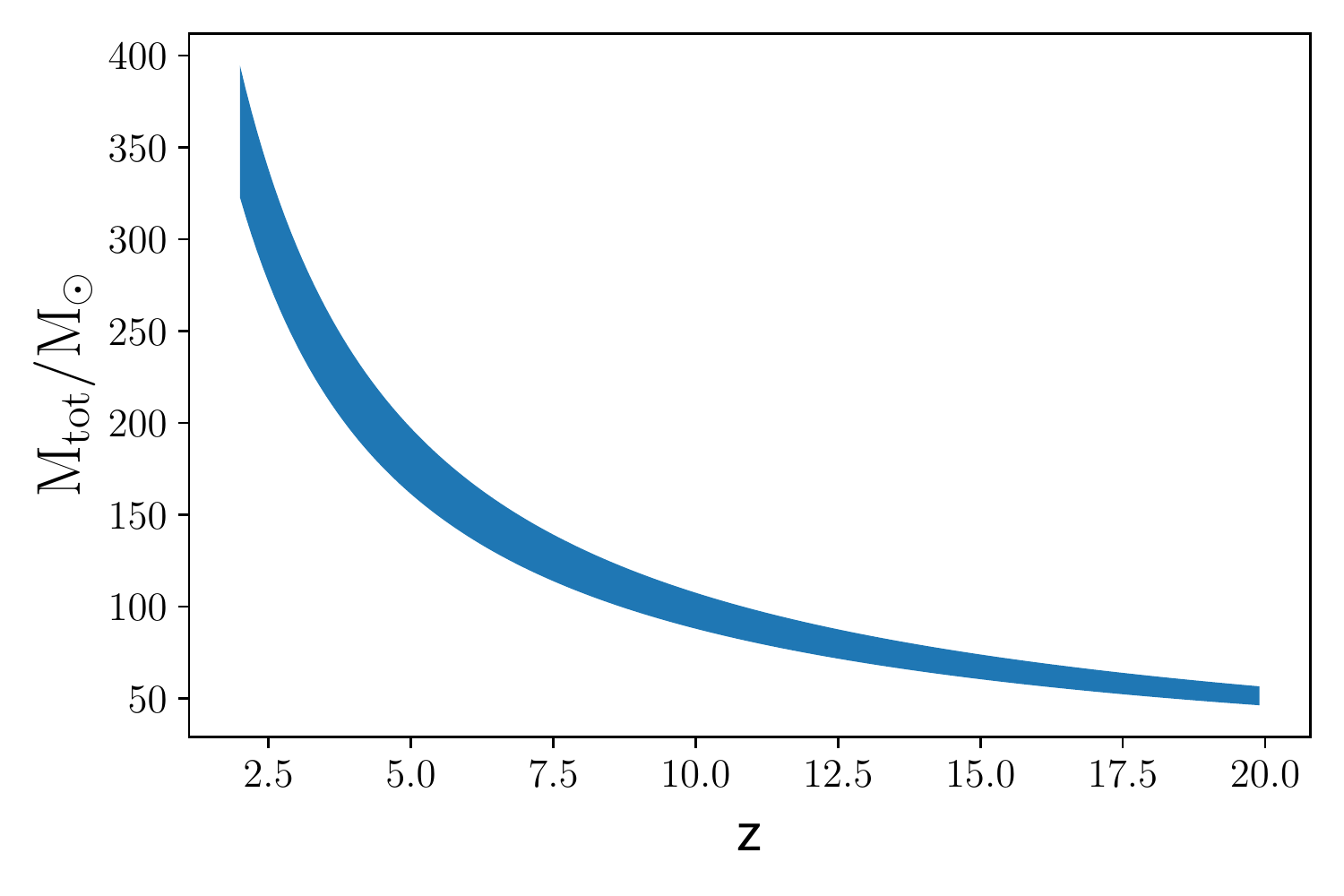}
    \caption{$M^z_{\rm tot} = 1076 M_{\odot}$ constraint shown in the $M_{\rm tot}$-z plane, including 10\% error bars accounting for the uncertainty of the estimate of $M^z_{\rm tot}$ from $f_{\rm peak}$.}
    \label{fig:M_z}
\end{figure}

Already with a varying-index PL search we deduce that the 2/3 filter is not appropriate, since the estimated power law index is $\alpha = -0.6$, as observed in the corner plot in Fig.~\ref{fig:corner_pl}. The more intricate filters, however, fit the $\Omega_{\rm GW}$ spectrum well and capture the presence of the peak successfully. In order to understand the redshifted mass distribution of pop III creating the GWB, we investigate the peak frequency of the signal. We obtain a good estimate of the peak frequency using a BPL search filter as shown in Fig.~\ref{fig:corner_bpl}, $f_{\rm peak} = 15.4$ Hz. The redshifted ringdown frequency that matches the peak frequency, $f^z_{\rm ring} = 15.4$ Hz, corresponds to $M^z_{\rm tot} = 1076 M_{\odot}$. The ST redshifted mass distribution shown in the top panel of Fig.~\ref{fig:distrib}, gives an average redshifted total mass of the residual population, $<M^z_{\rm tot}>=1121 M_{\odot}$. Therefore, our estimate of average $M^z_{\rm tot}$ agrees well with the true value. Finally, the estimated  $M^z_{\rm tot}$ can be depicted as a curve in the redshift-intrinsic total mass plane since $M^z_{\rm tot}=(1+z)M_{\rm tot}$, see Fig.~\ref{fig:M_z}. Note that we have included a 10\% uncertainty error in matching of the ringdown and the peak frequency for consistency with our findings in Table~\ref{tab:m_z}.

\textbf{Conclusions---}
GWs emitted from CBCs formed by pop III stars could represent a promising detection channel of the first stars formed in the Universe in a very low-metallicity environment. Although 2G detectors are unable to detect the GWB from pop III stars, we have demonstrated that future GW interferometers could lead to a detection. Employing the ET+2CE 3G detector network, we could successfully subtract enough individual merger events to unravel the elusive pop III GWB. Subtraction methods are more effective for nearby sources, so the residual pop I/II signal can become sub-dominant to the pop III residual. The 2/3 power law approximation of GWB energy density for the CBC background breaks in this case due to the higher redshifted masses of pop III. Instead of the 2/3 power law, the GWB spectrum peaks in the low-frequency LIGO/Virgo range.

A model selection study showed that we could observe the peak caused by the unresolved merger and ringdown GWs from pop III stars. With a good estimate of the peak frequency, we could even deduce the redshifted total mass distribution of the residual population. Taking data from the recent ST binary sources catalogue, we have demonstrated the effectiveness of our Bayesian analysis combined with implications of retrieved parameters. One should note that we have derived the relationship between peak frequency and redshifted total mass assuming equal-mass and non-spinning binaries. Future work should involve relaxing these assumptions. Additionally, one could further investigate the connection between the negative slope of the $\Omega_{\rm GW}$ spectrum and pop III properties such as merger rate and mass distribution.

\textbf{Acknowledgments---}The authors acknowledge access to computational resources provided by the LIGO Laboratory supported by National Science Foundation Grants PHY-0757058 and PHY-0823459. This paper has been given LIGO DCC number 2100337. K.M. is supported by King's College London through a Postgraduate International Scholarship. M.S. is supported in part by the Science and Technology Facility Council (STFC), United Kingdom, under the research grant ST/P000258/1. 

Numerous software packages were used in this paper. These include \texttt{matplotlib}~\cite{Hunter:2007}, \texttt{numpy}~\cite{numpy}, \texttt{scipy}~\cite{2020SciPy-NMeth}, \texttt{bilby}~\cite{Ashton:2018jfp}, \texttt{dynesty}~\cite{2020MNRAS.493.3132S}, \texttt{PyMultiNest}~\cite{refId0}.

\section*{Appendix}
\textbf{Priors---}All of the models have the same log-uniform prior distribution for the GWB amplitude ranging between $10^{-13}$ and $10^{-5}$. As for the remaining parameters, we use
\begin{itemize}
    \item PL: $\alpha=\mathcal{N}(0,3.5)$, 
    \item BPL: $\alpha_1=U(2/3,5/3), \alpha_2=U(0,-8), f_{\rm peak}=U(10,100)$, 
    \item smooth BPL: $\alpha_1=U(2/3,5/3)$, $\alpha_2=U(0,-8)$, $f_{\rm peak}=U(10,100), \Delta=U(0,10)$, 
    \item triple BPL: $\alpha_1=U(2/3,5/3)$, $\alpha_2=U(0,-8),  \alpha_3=\delta(2/3), f_{\rm peak}^1=U(10,100), f_{\rm peak}^2=U(10,100).$
\end{itemize}
For models with a break frequency we use a uniform prior for the first power law index between 2/3 and 5/3, since this represents the inspiral/merger regime of the CBC. Triple BPL has the third spectral index fixed to $\alpha_3=2/3$ since we expect the inspiral phase of pop I/II signal to dominate at higher frequencies.
\\

\begin{figure}
    \centering
    \includegraphics[width=0.5\textwidth]{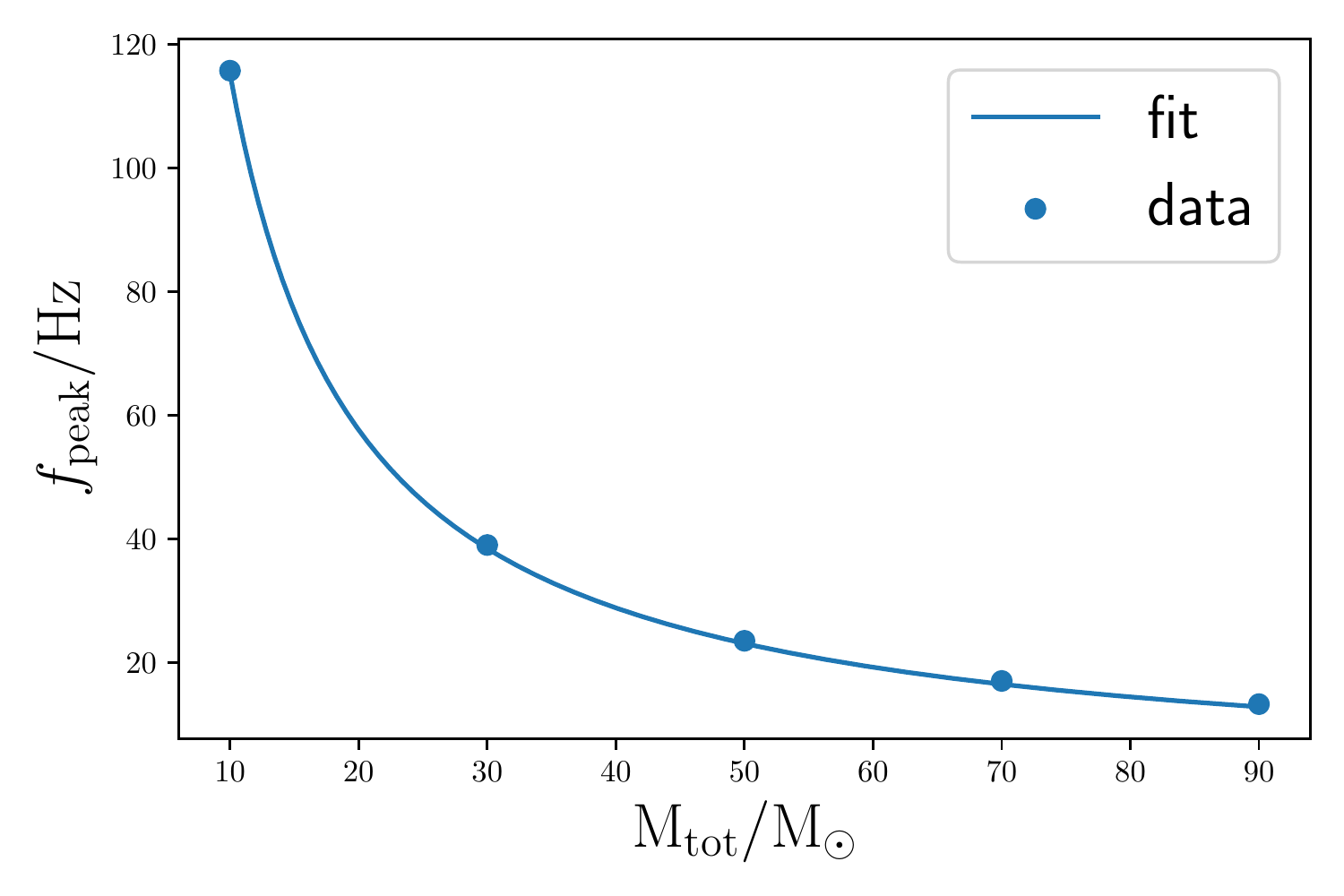}
    \caption{StarTrack merger rate evolved, equal-mass binaries. We find a relationship between peak frequency and total intrinsic mass of the merger. This is a model-dependent statement.}
    \label{fig:Mtot_fit}
\end{figure}

\textbf{Intrinsic mass distribution---}
We study the relation between peak frequency of pop III GWB spectrum and the mass distribution of the sources. We fix the merger rate as a function of redshift to be the one of ST. For total mass, $M_{\rm{tot}}=m_1+m_2$ varying between 10 and 90 $M_{\odot}$, we generate $\Omega_{\rm{GW}}$ spectra and record the frequency at which the spectra are maximum. We then find a best-fit curve for the data,

\[
f_{\rm peak} = f_0 \left(\frac{90 M_{\odot}}{M_{\rm tot}}\right) \rm Hz,
\]
with $f_0= 12.8$ Hz see Fig. \ref{fig:Mtot_fit}. However, changing the merger rate to the one from \cite{Inayoshi:2016hco}, we find a different best fit curve, with $f_0= 53.7$ Hz,
implying that the intrinsic mass may be difficult to extract from the estimate of the spectrum peak. We find more promising results if we study the redshifted total mass and its relation to peak frequency, as described in the main text.

\end{document}